\pdfoutput=1
\documentclass[letterpaper,preprint,fleqn,prd,superscriptaddress,nofootinbib,longbibliography]{revtex4-2}
\usepackage{amsfonts,amssymb,amsmath,mathtools,mathrsfs,slashed}	
\usepackage{graphicx}		
\usepackage{hyperref}		
\usepackage{array}
\usepackage[utf8]{inputenc}
\usepackage[shortlabels]{enumitem}
\usepackage{physics}
\usepackage{color}
\usepackage{comment}

\usepackage[top=2.5cm, bottom=2.5cm, left=2.5cm, right=2.5cm]{geometry}	

\def\R{{\mathcal{R}}}

\begin{document}

\title{Higgs phases at non-zero density from holography}

\author{Oscar Henriksson}
\email{oscar.henriksson@helsinki.fi}
\affiliation{Department of Physics and Helsinki Institute of Physics\\
P.O.~Box 64, FI-00014 University of Helsinki, Finland}

\author{Antti Hippel\"ainen}
\email{antti.hippelainen@helsinki.fi}
\affiliation{Department of Physics and Helsinki Institute of Physics\\
P.O.~Box 64, FI-00014 University of Helsinki, Finland}

\author{Carlos~Hoyos}
\email{hoyoscarlos@uniovi.es}
\affiliation{Department of Physics and \\
Instituto de Ciencias y Tecnolog\'{\i}as Espaciales de Asturias (ICTEA),\\
Universidad de Oviedo, c/ Federico Garc\'{\i}a Lorca 18, ES-33007 Oviedo, Spain}

\author{Niko Jokela}
\email{niko.jokela@helsinki.fi}
\affiliation{Department of Physics and Helsinki Institute of Physics\\
P.O.~Box 64, FI-00014 University of Helsinki, Finland}

\author{Aleksi Piispa}
\email{aleksi.piispa@helsinki.fi}
\affiliation{Department of Physics and Helsinki Institute of Physics\\
P.O.~Box 64, FI-00014 University of Helsinki, Finland}

\begin{abstract}
We show how Higgs phases at non-zero density can be described using a simple analytic method for gauge theories possessing a holographic dual. We introduce co-dimension one branes in a bottom-up gravity dual that are sources of form flux, such that the effective curvature radius is changed when the brane is crossed. This mimics the expected flow produced by color branes nucleating in a top-down model.
\end{abstract}

\preprint{HIP-2022-32/TH}

\maketitle

\newpage

\tableofcontents

\newpage

\section{Introduction}

Exploration of gauge theories and their phase diagrams continue to surprise us time and again. In this paper, we will study gauge theories at non-zero density and temperature, focusing on so-called Higgs phases. These phases can be described at weak coupling (but non-perturbatively) as the condensation of a scalar operator which breaks the gauge symmetry and gives a mass to (part of) the spectrum. As one important real-world example, we mention QCD; at asymptotically large baryon density, the theory can be treated as weakly coupled, and quark pairing can be shown to lead to a Higgs phase known as a color superconductor (see \cite{Alford:2007xm} for a review). However, at more moderate densities, such as those that might be realized in compact star cores, the coupling is no longer weak, and the precise fate of the color superconducting state is unknown.

Holographic duality has been successfully used to describe various phenomena and phases of strongly coupled gauge theories and one would expect it to be able to capture the physics of Higgs phases as well. However, this has proven a challenging task; previous attempts have either focused on bottom-up models describing it as global symmetry breaking \cite{Basu:2011yg,BitaghsirFadafan:2018iqr,Ghoroku:2019trx,Nam:2021ufk} or on a special top-down realization where the finite-density ground state is supersymmetric \cite{Faedo:2018fjw}.

Here we aim to improve on this situation by proposing a simple bottom-up scenario which can describe a class of holographic Higgs phases. The novel aspect of our scenario, which we detail in a specific case model, is the inclusion of explicit brane sources. This is how gauge symmetry breaking can be realized in well-understood top-down models as we now review.

\subsection{Top-down motivation}

Consider the familiar case of $\mathcal{N}=4$ super-Yang-Mills (SYM), holographically dual to type IIB string theory on $AdS_5\times S^5$. In the vacuum at zero temperature and vanishing R-charge density, this theory has a large moduli space corresponding to diagonal expectation values for the six adjoint scalar fields. In the gravity dual, this manifests through the BPS property of parallel D3-branes allowing them to be distributed arbitrarily in the six transverse directions at no cost in energy. While $N$ coincident branes gives rise to an SU($N$) gauge theory,\footnote{The theory of $N$ coincident D3-branes is actually U($N$), but in the near horizon limit that gives the holographic dual, the degrees of freedom of the diagonal Abelian group decouple.} a generic distribution of them breaks this to U(1)$^{N-1}$. Corresponding dual geometries in the large-$N$ limit can be constructed \cite{Kraus:1998hv,Freedman:1999gk,Lin:2004nb}.

Hence, in the vacuum, the existence of the moduli space makes it easy to find a Higgs phase in the theory.\footnote{Technically it is a Coulomb phase because an Abelian subgroup remains unbroken, but all the non-Abelian components become massive, so we will adopt a less rigorous but more physical terminology and refer to these phases as Higgsed.} The question is now, how does this change when turning on a temperature and, most importantly, an R-charge chemical potential?

At weak coupling, non-zero temperature gives a thermal mass to the massless scalars, while the chemical potential gives a negative contribution. As discussed in \cite{Basu:2005pj,Yamada:2006rx,Hollowood:2008gp}, at low temperatures and non-zero chemical potential this seems to lead to a fatal instability in the theory, giving an effective potential which is unbounded from below. In the holographic dual at strong coupling this picture was seemingly confirmed through a probe brane calculation: In \cite{Yamada:2008em} it was noted that the effective action of a single D3-brane is also unbounded from below in a black brane geometry corresponding to a finite-density deconfined state \cite{Cvetic:1999xp} .

This was, however, re-examined in \cite{Henriksson:2019zph}, which came to a somewhat different conclusion. The black brane backgrounds in question rotate in the $S^5$-part of the geometry, the angular momentum being dual to the R-charge density of the field theory. While \cite{Yamada:2008em} assumed that the D3-brane probe rotates at exactly the angular velocity of the event horizon, \cite{Henriksson:2019zph} instead took the angular momentum of the brane to be constant, as required by the equations of motion. It was then observed that there is a critical angular momentum for the brane to signal an instability. Furthermore, even for supercritical values of the angular momentum, the potential remains bounded from below, but now possessing a global minimum away from the horizon. When the field theory lives in flat space, this minimum is infinitely far away, at the boundary of AdS. Interestingly, however, when the field theory lives on a sphere, so that the dual geometry is global AdS, the global minimum is at a finite position in the holographic radial direction. In that case, it was argued that a (meta)stable Higgs phase will result, in general with some fraction of the branes localized in the bulk and some behind the horizon. If $M$ out of the $N$ branes nucleate in the bulk, the gauge symmetry will be effectively broken as $SU(N)\to SU(N-M)\times U(1)^M$.

To determine exactly which is the final state of this instability for a given temperature and chemical potential, one must take the backreaction of the nucleated branes into account. This is a primary motivation that sparked our present work. As an initial step towards constructing Higgs phases in the holographic dual of ${\mathcal{N}}=4$ SYM, we will consider a 5D bottom-up model that captures its main features.

In more detail, we will include explicit 3-brane sources, which capture all the relevant features of the full top-down construction. This approach has two benefits. First, it is somewhat simpler to treat, since we do not need to detail how the internal geometry is affected. At any rate, one can consistently truncate the ten-dimensional geometry and would presumably end up with a setup very similar to the case at hand. Second, it emphasizes the generality of the phenomena we study; this scenario can likely be implemented in a variety of holographic settings.  In fact, in addition to extending our case study to the dual of $\mathcal{N}=4$ SYM, it is worth pointing out that other top-down theories, such as the Klebanov-Witten gauge theory \cite{Klebanov:1998hh}, share similar brane nucleation instabilities  \cite{Herzog:2009gd,Henriksson:2019ifu,Henriksson:2021zei} and will likely therefore also have Higgs phases. We thus believe that our program will provide valuable lessons on understanding the properties of Higgs phases of strongly coupled gauge theories in general.

The rest of the paper is organized as follows. Section~\ref{sec:model} introduces and further motivates our setup: a five-dimensional gravity theory with a one-form gauge field (holographically dual to a conserved current), a four-form gauge field, and 3-branes which carry charge under both the one-form and four-form fields. In Section~\ref{sec:junction} we use Israel junction conditions to find static solutions with 3-brane sources forming a thin domain wall, or a shell, localized in the holographic radial coordinate. Section~\ref{sec:thermo} provides a study of the thermodynamics of these shell solutions and the resulting phase diagram, including the phase transition between shell solutions and regular AdS-Reissner-Nordström (AdS-RN). We conclude and discuss future directions in Section~\ref{sec:discussion}. In Appendix~\ref{app:10Dprobemodel} we review parts of the probe D3-brane computations from \cite{Henriksson:2019zph,Danielsson:2022lsl} which motivates our approach.

\section{Bottom-up gravity model}\label{sec:model}

We aim to construct the simplest possible holographic model that could capture the physics of a finite-density Higgs phase. We thus start from a 5D gravitational theory with a negative cosmological constant and a gauge field, with two-form field strength $F_2=dA_1$, which will be dual to a conserved current. To this workhorse of applied holography we add a five-form field strength $F_5=dC_4$, giving the following bulk action:
\begin{equation}\label{eq:bulkaction}
S_\text{bulk}=\frac{1}{2\kappa^2}\int d^{5}x\sqrt{-g}\left( \R+\frac{12}{L^{2}}-L^2 F_2^2-\frac{1}{2\cdot 5!} F_5^2\right) + \frac{1}{4!}\int d^5x\, \partial_{\mu_1} \left( \sqrt{-g}F^{\mu_1\ldots\mu_5}C_{\mu_2\ldots\mu_5}  \right) \ .
\end{equation}
Here, $\R$ is the Ricci scalar, $L$ is the radius of curvature, and $\kappa^2=8\pi G_5$ with $G_5$ the Newton constant in five dimensions. 

The addition of $F_5$ is motivated by string theory constructions where the rank of the group is linked to the flux of the form potential. This potential couples to the color branes that source the geometry and eventually determines the radius of curvature. If the field theory is in a Higgs phase, we expect the effective rank to change as we move from the UV (near the AdS boundary) to the IR (deep inside the AdS bulk). This then indicates that we should allow the effective AdS radius to vary. The five-form field strength makes this possible, as described by Brown and Teitelboim \cite{Brown:1987dd,Brown:1988kg}. In the absence of sources, the equations of motion set it equal to
\begin{equation}
    F_{\mu_1\ldots\mu_5}=f\sqrt{-g}\, \epsilon_{\mu_1\ldots\mu_5} \ ,
\end{equation}
with $f$ a constant. Substituting this back into the bulk action gives
\begin{equation}
S_\text{bulk}=\frac{1}{2\kappa^2}\int d^{5}x\sqrt{-g}\left( \R+\frac{12}{L^{2}}-\frac{1}{2} f^2-L^2 F_2^2\right) \ ,
\end{equation}
so we see that the flux of $F_5$ shifts the effective cosmological constant to\footnote{The boundary term in \eqref{eq:bulkaction} was chosen such that the Einstein equations agree with this value.}
\begin{equation}\label{eq:shiftedLambda}
    \Lambda=-\frac{6}{L_\text{eff}^2}=-\frac{6}{L^{2}}+\frac{1}{4} f^2 \ .
\end{equation}

We will identify the cosmological constant in the absence of fluxes as the value that corresponds to the un-Higgsed phase, which should thus be the value at the AdS boundary. We therefore impose the condition that the flux vanishes at the boundary. Moving to the interior from the boundary can result to a non-vanishing flux which reduces the magnitude of the effective cosmological constant. The maximum magnitude for the flux $f=\pm f_\text{max}$, corresponding to a completely Higgsed phase, is the one that makes the effective cosmological constant vanishing
\begin{equation}      \label{eq:fmax}
f_\text{max}=\frac{2\sqrt{6}}{L}\ .
\end{equation}

In order to describe the change in the flux through the geometry we furthermore add 3-branes to our theory, with an action given by
\begin{equation}\label{eq:braneAction}
    S_{\text{brane}}=-T_3\int d^4\xi \sqrt{-\det P[g]} + \mu_3 \int P[C_4] + \int P[A_1]\wedge J_3 \ .
\end{equation}
The total action is then $S=S_{\text{bulk}}+S_{\text{brane}}$. The branes are codimension-one and thus act as domain walls in our 5D spacetime. Their action is clearly inspired by the action of a D3-brane in string theory: The first term is of the standard Dirac-Born-Infeld (DBI) form, with $T_3$ the tension and $P[g]$ the pullback of the 5D metric $g_{\mu\nu}$. The second term is a standard Wess-Zumino (WZ) coupling to the potential of the five-form $F_5=dC_4$, with charge $\mu_3$. This means that the flux of $F_5$, and the effective cosmological constant, will change upon crossing a 3-brane domain wall. The difference in fluxes between the two sides of the wall is
\begin{equation}\label{eq:fluxjump}
    \Delta f=2\kappa^2 \mu_3\,. 
\end{equation}
Therefore, from \eqref{eq:shiftedLambda}, the change in the cosmological constant is
\begin{equation}\label{eq:lambdajump}
    \Delta \Lambda=\kappa^2\mu_3\bar{f}\,,
\end{equation}
where $\bar{f}$ is the average value $\bar{f}=(f_1+f_2)/2$.

Note that, while in a top-down scenario, the entirety of the cosmological constant is typically due to the branes, in our bottom-up framework we allow ourselves to set its ``bare'' value (corresponding to $L$) as we see fit. Moreover, we are free to set the value of the five-form flux near the AdS boundary, as a boundary condition. In a top-down realization, these will be related.

The third term in (\ref{eq:braneAction}) does not appear in the standard D-brane action; it introduces a coupling to the 5D gauge field $A_1$ (with $F_2=dA_1$) through a worldvolume three-form $J_3$, which must be closed, $dJ_3=0$, to respect gauge invariance. The idea that the 3-branes should couple to $P[A_1]$ in this way is again motivated by top-down computations in, {\emph{e.g.}}, the dual of $\mathcal{N}=4$ SYM at non-zero density. While the fundamental D-brane action does not have such a coupling, it was seen in \cite{Henriksson:2019zph,Danielsson:2022lsl} that a similar term is effectively induced through the embedding in the dual spinning black brane geometry. As we review in the appendix, this follows from the fact that for the brane to nucleate, it needs some non-zero angular momentum; the angular momentum in turn couples to off-diagonal components of the metric which, upon reduction to five dimensions, act as the gauge field(s) dual to the R-charge current. The physical picture to have in mind is that larger charge densities lead to a repulsion between the branes, which will lead to them separating and thus Higgsing the dual theory.

We expect that such a picture will continue to hold true in many other top-down instances of holography, possibly with modified couplings between the gauge field and the branes. Thus, this setup can lead to interesting possibilities for holographic model building with the ability to adapt both the bulk as well as the brane actions to describe the physics of interest.

\subsection{Regular black hole solutions}

In the absence of 3-branes, the five-form flux can be set to zero, and our theory has standard AdS-RN charged black hole solutions,
\begin{equation}
ds^{2}=-f(r)dt^{2}+f(r)^{-1}dr^{2}+r^{2}d\Omega_{3}^{2}
\end{equation}
with
\begin{equation} \label{eq:f}
f(r)=\frac{r^{2}}{L^{2}}+1-\frac{m}{r^{2}}+\frac{q^{2}}{r^{4}} \ ,
\end{equation}
and with $d\Omega_3^2$ being the metric of a unit three-sphere. The event horizon radius can be defined as the largest root of $f(r_H)=0$. The existence of an event horizon imposes a bound on the parameters, which can be written in the form $|q|\ge q_\text{ext}(m)$ with $q_\text{ext}(m)$ a fairly complicated function which we will not reproduce here; if this inequality is saturated, the black hole is extremal. Note, however, that if the theory admits a supersymmetric embedding, such extremal black holes are not BPS \cite{Chamblin:1999tk}.

The corresponding one-form gauge field, which we require to approach zero at the horizon, is 
\begin{equation}\label{eq:GaugeField}
 A_1=\frac{\sqrt{3}q}{2L}\left(\frac{1}{r_H^2}-\frac{1}{r^{2}}\right)dt\ .
\end{equation}
All of the solutions we study will asymptote to global AdS, so the dual field theory lives on a three-sphere, whose radius is (for now) set to equal $L$.

\section{Thin-shell solutions}\label{sec:junction}

The model introduced in the previous section can in principle be solved for arbitrary distributions of 3-branes in the holographic radial direction. Here, we will focus on a simple special case, by assuming that

\begin{enumerate}[(i),topsep=-1pt,itemsep=0ex,partopsep=3ex,parsep=2ex]
    \item the nucleated branes are tightly distributed around a certain radius, meaning we can approximate them by a thin shell partitioning the 5D spacetime into an interior and an exterior, and
    \item that \emph{all} the branes have nucleated, meaning that the interior spacetime of the shell is completely sourceless, a flat Minkowski.
\end{enumerate}
The resulting geometry is analogous to the vacuum Coulomb-branch solution studied in \cite{Kraus:1998hv}, where $N$ D3-branes are distributed in an $SO(6)$-symmetric shell with a flat interior.

With this configuration, the five-form flux in the interior of the geometry can be taken to be $f=-f_\text{max}$ in \eqref{eq:fmax}, while the flux outside vanishes. This implies that $N$ times $\Delta f$ in \eqref{eq:fluxjump} equals $f_\text{max}$, which fixes the relation between the number of branes and the asymptotic AdS radius
\begin{equation}  \label{eq:mu3val}
     L=\frac{\sqrt{6}}{N\kappa^2 \mu_3}\ .
\end{equation}
Let us momentarily assume that our model corresponds to a dimensional reduction of a ten-dimensional string construction with a geometry asymptotically $AdS_5\times {\cal M}_5$, and that the domain walls actually correspond to D3-branes. In this case
\begin{equation}
\kappa^2=\frac{\kappa_{10}^2}{\text{Vol}({\cal M}_5)}\ , \ \kappa_{10}^2=\frac{1}{2}(2\pi)^7g_s^2(\alpha')^4\ , \  \mu_3=\frac{g_3}{(2\pi)^3g_s(\alpha')^2}\ ,
\end{equation}
where $g_3$ takes into account that the normalization of the five-form in the reduced theory might be different from the normalization in ten dimensions. The volume of the internal space is expected to have a size of the same order as the AdS radius, so that $\text{Vol}({\cal M}_5)=v_5 L^5$. Together with (\ref{eq:mu3val}) this leads to relations of the form
\begin{equation}
    L^4/(\alpha')^2= \frac{4\pi^4}{v_5}\sqrt{\frac{2}{3}}g_3 g_s N \ , \  \ \ \frac{L^8}{G_{10}}=\frac{4\pi^2}{3 v_5^2}g_3^2 N^2\ ,
\end{equation}
which is similar to what is found in the ten-dimensional theory. For instance, one can recover the same expressions as the $AdS_5\times S^5$ dual to ${\cal N}=4$ super Yang-Mills by setting $v_5=\pi^3/3$ and $g_3=1/\sqrt{6}$.

While we believe that on a generic point of the phase diagram only a fraction of all the branes will nucleate, it seems plausible that there would exist regions where all, or almost all, branes nucleate, motivating assumption \textit{(ii)}. Assumption \textit{(i)} can be motivated as follows: In the top-down treatment of \cite{Henriksson:2019zph}, it was argued that the $N$ D3-branes will tend to share the total charge (in that case, angular momentum) equally. From the probe computations in that paper, one then concludes that they all prefer to localize around the same radius.

Assumption \textit{(i)} lets us treat the problem using Israel junction conditions \cite{Israel:1966rt}; on each side of the shell, the solution will be of the general AdS-RN form discussed in the previous section, but the parameters $m$, $q$, and $L_{\text{eff}}$ will differ on the two sides. We use $+$/$-$ to denote quantities on the outside/inside of the shell. Then, by assumption \textit{(ii)}, inside the shell we take $f^-(r)=1$. Outside we allow for a general blackening function $f^+(r)=f(r)$; we take as boundary condition that the flux $f$ is zero, allowing us to replace $L_\text{eff}$ with $L$. The shell hypersurface will be parameterized by the angular coordinates $(\psi,\theta,\phi)$ on the spacetime three-sphere it wraps, together with the time coordinate of the inner, flat solution $t^-\equiv\tau$. The shell will be localized in the radial direction at $r^\pm=R^\pm$.

The first junction condition states that the induced metric on the shell should be the same when approached from both directions. With said parametrization, it becomes
\begin{equation}
ds_{B}^{2}=h_{ab}d\xi^a d\xi^b = -(dt^-)^{2}+(R^-)^2d\Omega_{3}^{2}=-f(R^+)(dt^+)^2+(R^+)^2d\Omega_{3}^{2} \ .
\end{equation}
Equality of the spatial part of this metric means that the radial coordinate of the shell is the same in both the coordinates of the interior and exterior solution, $R^+=R^-\equiv R$. The time coordinate, however, has a discontinuity on the shell, $\sqrt{f(R)}t^+ = t^-$.

The second junction condition is
\begin{equation}\label{eq:2ndJunctionCond}
 K_{ab}^{+}-K_{ab}^{-}=-\kappa^2\Bigg(T_{ab}-\frac{1}{3}T_{c}^{c}h_{ab}\Bigg) \ ,
\end{equation}
where $K_{ab}$ is the extrinsic curvature and $T_{ab}$ is the shell stress tensor. The extrinsic curvature can be computed using
\begin{equation}
    K_{ab}= -n_\mu \left( \frac{\partial^2 X^\mu}{\partial\xi^a \partial\xi^b} + \Gamma^\mu_{\nu\lambda} \frac{\partial X^\nu}{\partial\xi^a}\frac{\partial X^\lambda}{\partial\xi^b} \right) \ ,
\end{equation}
where $n^\mu$ is the unit normal of the shell, $X^\mu$ is the embedding of the shell, and $\Gamma^\mu_{\nu\lambda}$ are the Christoffel symbols. For the exterior AdS-RN geometry, this gives the result
\begin{equation}
    K^\tau_\tau = \frac{f'(R)}{2\sqrt{f(R)}} \ , \qquad K^\psi_\psi=K^\theta_\theta=K^\phi_\phi = \frac{\sqrt{f(R)}}{R} \ ,
\end{equation}
while the results for the interior flat geometry are the same with $f(R)=1$.

Since the shell is supposed to consist of $N$ 3-branes, the shell stress tensor can be computed from the variation of (\ref{eq:braneAction}) with respect to the induced metric. Only the DBI-term contributes to this variation, giving a worldvolume cosmological constant,
\begin{equation}
    T_{ab} = -\frac{2}{\sqrt{-h}}\frac{\delta S_\text{brane}}{\delta h^{ab}} = -N T_3 h_{ab} \ .
\end{equation}

To proceed we must choose a value for the tension $T_3$ of our 3-branes. We do this by, again, mimicking the top-down probe computation of \cite{Henriksson:2019zph}; from the effective potential therein one finds that the D3-brane tension is \emph{critical}, meaning that the effective worldvolume cosmological constant which receives contributions from $T_3$ as well as from the embedding is zero.\footnote{For a more detailed discussion of this see \cite{Danielsson:2022lsl}, or \cite{Kraus:1999it} in a Randall-Sundrum setting (where the critical tension differs from here by a factor of 2).} This corresponds to a tension
\begin{equation}
    T_3=\frac{3}{\kappa^2 N L} =\sqrt{\frac{3}{2}}\, \mu_3\ ,
\end{equation}
where we have used \eqref{eq:mu3val} in the last equality. Inserting the extrinsic curvature for the flat and AdS-RN spacetimes, we now obtain two independent equations,
\begin{equation}
\label{eq:2ndJunctionCondExpl}
-\frac{f'(R)}{2\sqrt{f(R)}} = \frac{1}{L} \qquad \text{and} \qquad \sqrt{f(R)}-1 =-\frac{R}{L} \ ,
\end{equation}
coming from the time-time and the space-space components of (\ref{eq:2ndJunctionCond}), respectively. These equations are solved by
\begin{equation}\label{eq:shellSol}
    m=\frac{5R^3}{L} \qquad \text{and} \qquad q=\pm\sqrt{\frac{3R^5}{L}} \ ,
\end{equation}
\emph{provided} $0\le R\le L$. We thus have a one-parameter family of static shell solutions. 

\begin{figure}[ht]
 \centering
 \includegraphics[scale=1.]{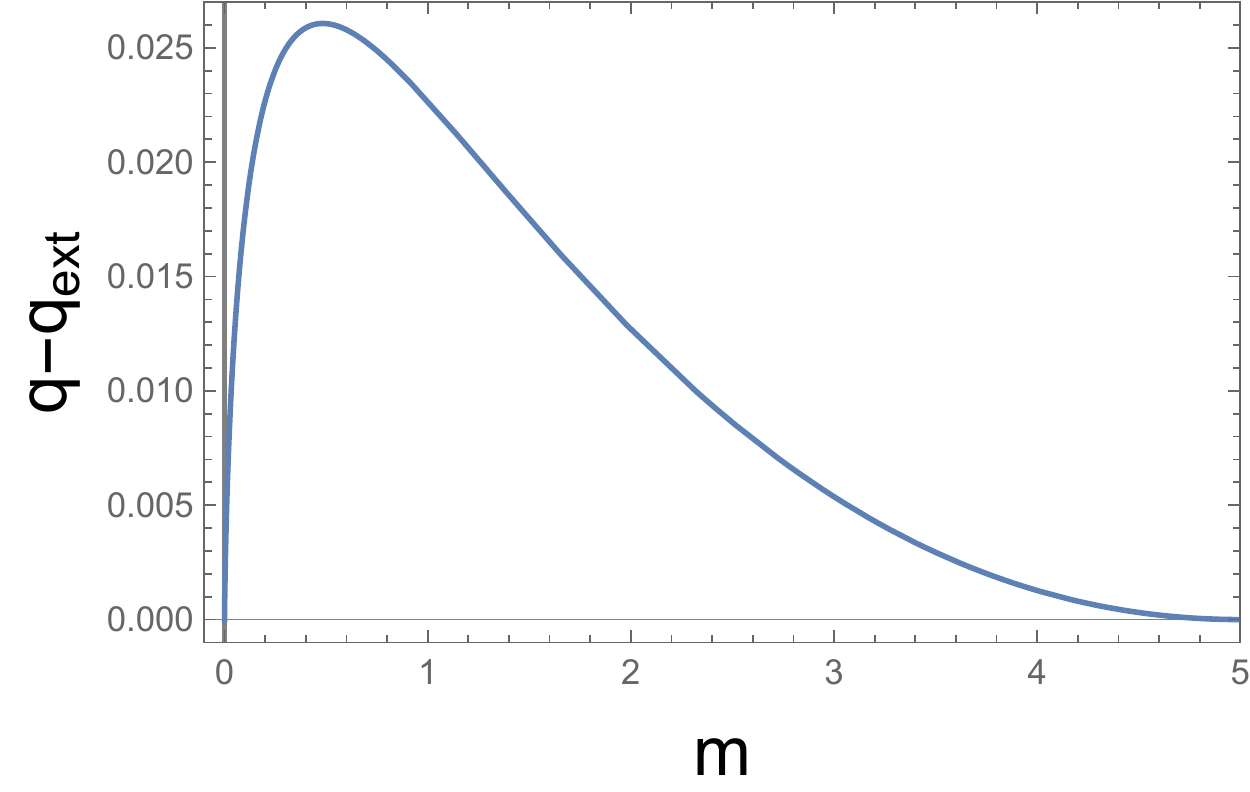}
 \caption{The difference between the charge of the shell solution (\ref{eq:shellSol}) and the extremal charge $q_\text{ext}$ as a function of mass, in units of $L$.}\label{fig:overExtremal}
\end{figure}

It turns out that the solution (\ref{eq:shellSol}) corresponds to an over-extremal black hole. In Fig.~\ref{fig:overExtremal} this is visualized by plotting the charge of the shell solution minus the extremal charge, as a function of mass (in units of $L$). The charge is seen to be over-extremal everywhere except at $m=0$ (no black hole) and at $m/L^2=5$, corresponding to $R=L$; at this special point the exterior AdS-RN geometry in the shell solution becomes extremal, with the shell position coinciding with the horizon. Everywhere else, the exterior geometry of the shell solutions has a naked singularity; however, when the region interior to the shell is replaced by flat space, the singularity is excised. This is somewhat reminiscent of other stringy resolutions of singularities such as the enhan\c con \cite{Johnson:1999qt}. In fact, it has been argued in a top-down setting \cite{Evans:2002fk} that certain solutions with naked singularities can be relevant thermodynamic phases, which can even be preferred in some parts of the phase diagram (that reference focused on the canonical ensemble). If our shell solutions can be uplifted to such top-down settings, it is plausible that they can replace such naked singularities in the phase diagram.

The fact that the exterior solution typically has a naked singularity means that the shell will never risk falling into a horizon. Thus we can also consider the limit $R\to 0$; we see that this also brings $m$ and $q$ to zero, meaning that the exterior geometry, which is now all there is, approaches thermal AdS.

\section{Thermodynamics and phase diagram}
\label{sec:thermo}

Before studying the thermodynamics of the shell solutions, we review the thermodynamics of the regular AdS-RN solutions. This can be obtained by standard methods and is found in many references, see for instance \cite{Chamblin:1999tk}. 

Note that up to now, we have worked in units in which the 3-sphere radius on which the field theory is defined has been set equal to $L_{\text{eff}}$. In this section, we will allow for an arbitrary radius $\rho$. This is accomplished by rescaling the coordinates $r\to(\rho/L)r$, $t\to(L/\rho)t$ with a simultaneous rescaling of the parameters $m\to\frac{\rho^{4}}{L^{4}}m$, $q\to\frac{\rho^{3}}{L^{3}}q$. At large $r$ the metric now assumes the canonical AdS form
\begin{equation}
ds^{2}\to -\frac{r^{2}}{L^{2}}dt^{2}+\frac{L^{2}}{r^{2}}dr^{2}+\frac{\rho^{2}}{L^{2}}r^{2}d\Omega_{3}^{2} \ .
\end{equation}
For shell solutions, we only rescale the coordinates of the exterior AdS-RN solution (so after the rescaling, the radial coordinate is no longer continuous at the shell). The volume of the three-sphere on which the field theory lives is then $V=2\pi^2 \rho^3$. The stress tensor can be determined by evaluating the Brown-York tensor at the AdS-RN boundary with the standard counterterms. The resulting energy $E$ and pressure $P$ are
\begin{equation}
    E=3VP=\frac{3\pi^2 \rho^3}{\kappa^2 L^3}m \ .
\end{equation}
The entropy is given by the horizon area, and the temperature is obtained by imposing regularity near the horizon in imaginary time:
\begin{equation}
    S=\frac{4\pi^3 \rho^3}{\kappa^2 L^3}r_H^3 \ , \qquad T=\frac{f'(r=r_H)}{4\pi} \ .
\end{equation}
Finally, the conserved charge and the chemical potential are
\begin{equation}
    Q=\frac{4\sqrt{3}\pi^2 \rho^3}{\kappa^2 L^2}q \ , \qquad \mu=\frac{\sqrt{3}q}{2L r_H^2} \ .
\end{equation}
These expressions can be seen to obey the first law of thermodynamics, $dE=TdS+\mu dQ-PdV$. The free energy can be computed (in the grand canonical ensemble) as $\Omega=E-TS-\mu Q$.

\subsection{Thermodynamics of shell solutions}

The thermodynamics for our shell solutions is obtained by similar methods. In particular, the expressions for energy, pressure, and charge density are the same as above, with the parameters $m$, $q$, and $L$ being those of the exterior AdS-RN solution.

The shell solutions are flat inside, and in the absence of a horizon have vanishing entropy to leading order in the large-$N$ expansion. For the same reason, the temperature of a particular shell solution is unconstrained and can be chosen at will.

Naively, the chemical potential is unconstrained in a similar way. It is dual to the asymptotic value of $A_t$; for regular AdS-RN solutions, the gauge field should be set to zero on the horizon, but in the absence of a horizon it can be shifted by an arbitrary constant. However, we argue that there must be a relationship between the chemical potential and the shell position, in order to satisfy the first law in the presence of a non-trivial charge density. (It is clear that, for non-zero $dQ$, the first law cannot be satisfied for arbitrary $\mu$.) This requirement gives
\begin{equation}
    \mu_s=\frac{\sqrt{3}q}{2LR^2} = \frac{3}{2L}\sqrt{\frac{R}{\rho}} \ ,
\end{equation}
which means that the shell is forced to sit at the radius where the gauge field vanishes. The same result can be found by considering an inside geometry with a black hole of mass $\tilde m$ and taking the limit $\tilde m \to 0$.

With these relations, we again find (by construction) that the first law is satisfied, and we can compute the free energy of our shell solutions, with the simple result
\begin{equation}
    \Omega_s = -\frac{3\pi^2 \rho^2 R^3}{\kappa^2 L^3} = -\frac{64\pi^2L^3\rho^5\mu^6}{243\kappa^2} \ .
\end{equation}
Recall that the shell solutions only existed for $R\le L$, or $R\le L^2/\rho$ in the rescaled units used in this section. This corresponds to chemical potentials $\rho\mu\le 3/2$. The point $R=L^2/\rho$, $\rho\mu=3/2$ is also the sole special point discussed in the previous section where the blackening function $f(r)$ has a (double) zero and the shell sits at the position of the extremal horizon.

It might seem surprising that our proposed Higgs phase exists only below a certain value of the chemical potential. However, note that we have restricted ourselves to studying a particularly simple version of the Higgs phase, with all branes having nucleated, leaving an empty interior. If one allows for more general interiors, describing partial nucleation with black holes inside the shell, it is plausible that one will find solutions at larger densities as well.

\subsection{Phase diagram}

In the absence of the shell solutions, there are two standard geometries, AdS-RN and thermal AdS. Thermal AdS has trivial thermodynamics at leading order in the large-$N$ expansion, with zero energy, entropy, and charge density (we are subtracting off the Casimir energy throughout). The results for AdS-RN were covered above. It can be shown that for small temperatures and chemical potentials, in units of the three-sphere radius, no black hole solutions exist and thermal AdS is the prevailing phase, while at larger temperatures and/or chemical potentials, AdS-RN dominates. At $\mu=0$, this corresponds to the famous Hawking-Page transition \cite{Hawking:1982dh}. Note that there are two branches of black hole solutions; in the grand canonical ensemble which we focus on here, only one of these (the ``large black holes'') are relevant for the phase diagram.
In the canonical or microcanonical ensemble, the small black hole phases can dominate over the large ones, leading to an interesting interpretation on the phase structure of the dual field theory \cite{Chamblin:1999tk,Asplund:2008xd,Jokela:2015sza,Hanada:2022wcq}.

\begin{figure}
 \centering
 \includegraphics[scale=0.75]{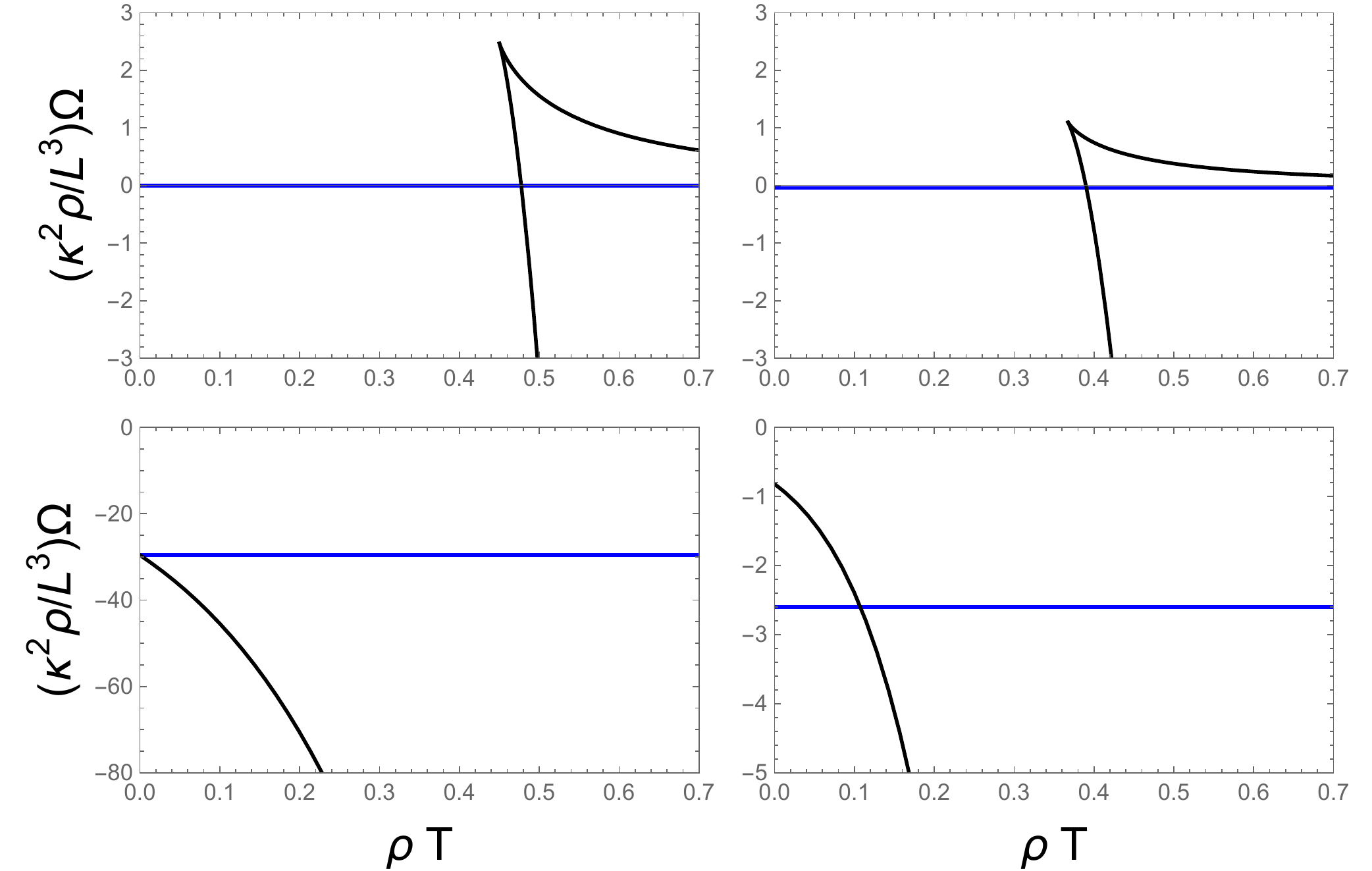}
 \caption{Free energy of the shell solutions (blue curve) and AdS-RN (black curves) as a function of $T$ at fixed $\mu$, in units of the three-sphere radius $\rho$. The chemical potentials are (clockwise from the top left) $\rho\mu=(0,0.5,1,1.5)$. For small chemical potentials (upper plots) there are two branches of black hole solutions (upper branch corresponding to the thermodynamically subdominant small BH solutions). Above $\rho\mu=1.5$ (bottom left) no shell solution exists.}\label{fig:freeEnergy}
\end{figure}

Turning to the shell solutions, we note that their free energy is always of order $N^2$ and negative for non-zero $\mu$, hence they will always dominate over thermal AdS (in the limit $\mu\to0$ the shell solutions reduce to thermal AdS, as previously discussed). Comparing their free energy with that of AdS-RN we arrive at Fig.~\ref{fig:freeEnergy}. We observe a first order phase transition with the shells being dominant at lower temperatures. The critical temperature approaches zero as $\rho\mu$ approaches 3/2. Note that at chemical potentials below $\rho\mu=\sqrt{3}/2$ there are two branches of black hole solutions; the branch with higher free energy corresponds to black holes with smaller radii which are thermodynamically subdominant.

\begin{figure}
 \centering
 \includegraphics[scale=1.]{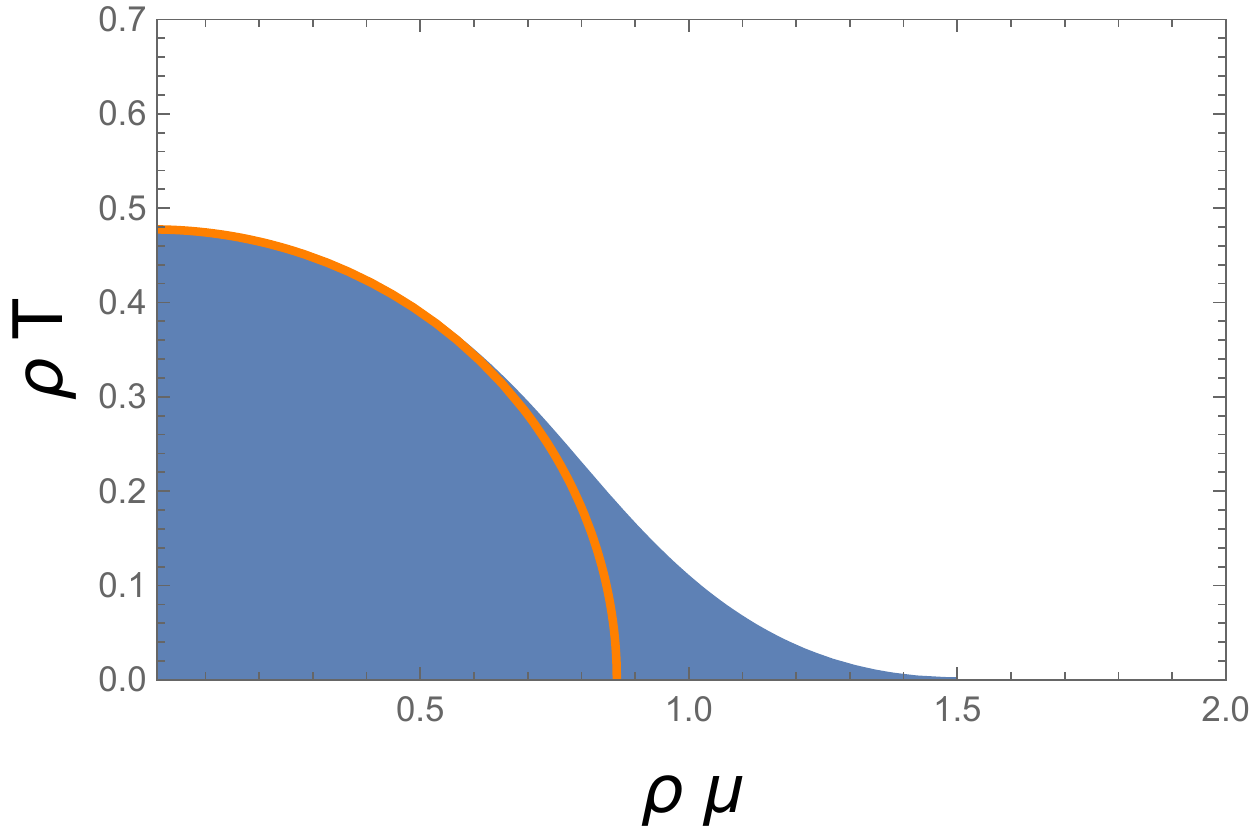}
 \caption{The phase diagram, with temperature $T$ and chemical potential $\mu$ in units of the three-sphere radius $\rho$. The white region shows where the normal, deconfined plasma described by AdS-RN dominates. The blue region denotes where the shell solution dominates. The orange curve denotes where the phase transition between AdS-RN and thermal AdS would be in the \emph{absence} of the shell solution.}\label{fig:phaseDiagram}
\end{figure}

The resulting phase diagram is shown in Fig.~\ref{fig:phaseDiagram}. The shell solutions are preferred in the blue region, AdS-RN is preferred in the white region, and the orange curve indicates where the phase transition to thermal AdS would be in the absence of the shell solutions. We note that the line of phase transitions between the shell solutions and AdS-RN follows the orange curve closely at small $\mu$; it can be shown that it asymptotes to the orange curve from above, with their difference going as $\mu^6$. This agrees with what was stated earlier; in the limit $\mu_s\to 0$ (equivalent to $R\to 0$), the shell solutions approach thermal AdS.

\section{Discussion}\label{sec:discussion}

In this paper we have proposed a simple holographic bottom-up model for finite-density Higgs phases akin to color superconductivity in QCD. Motivated by top-down considerations, explicit brane sources played an important role; the Higgs phase is characterized by the appearance --- or \emph{nucleation} --- of branes in the bulk of the gravity solution, outside possible event horizons.

We moreover considered a simple class of novel solutions to our model, describing a thin shell of branes partitioning spacetime into an interior and an exterior; we assumed a ``fully nucleated'' state, meaning the interior is source free and thus flat, while the exterior is described by an AdS-RN geometry. Using the Israel junction conditions we found a family of static shells, studied their thermodynamic properties, and verified that they obey the first law of thermodynamics. We compared their free energy (in the grand canonical ensemble) to that of the standard AdS-RN solution, as well as thermal AdS, and found that there exists a region at small $T$ and $\mu$ where the shell solution dominates. This region (blue in Fig.~\ref{fig:phaseDiagram}) completely overwhelms the thermal AdS, but is also the preferred phase over the AdS-RN for slightly larger chemical potentials. The phase transition between AdS-RN and the shell solutions is everywhere first order.

Let us examine the state corresponding to the simple ``shell phase'' solutions in some more detail. The free energy of the corresponding state scales as $N^2$, thus it is unlike the ``confined'' thermal AdS phase, which has zero free energy at this order.\footnote{We have in mind here ``kinematic confinement'' which forbids physical colorful states on the sphere, {\emph{i.e.}}, Gauss law disallowing charged states on compact space.} To probe the nature of the state in more detail could be exposed by loop operators. While we have been studying field theory on a three-sphere, we can borrow some intuition from a field theory in flat space. Wilson and 't Hooft loops can be computed from the actions of fundamental or D-strings, respectively. Following the reasoning in \cite{Rozali:2012ry}, we then expect shell solutions like ours to be classified as Higgs phases, in particular, displaying an area law for the 't Hooft loop at least when the radius of the three-sphere is very large.

Another class of probes of the properties of Higgs phases are two-point functions of gauge-invariant operators. Known supergravity solutions dual to the ${\cal N}=4$ SYM Coulomb branch predict a gapped spectrum with the exception of a massless mode corresponding to the dilaton of the spontaneously broken conformal invariance \cite{Freedman:1999gk,Bianchi:2001de}. One might expect to find similar results for the shell solutions.

The fully nucleated solution is a simple special case of more general shell solutions, where the interior of the shell would be an AdS-RN geometry with arbitrary values for $m$, $q$, and $L$. Then, one might anticipate to find several possible shell solutions at various points in the phase diagram and one must minimize the free energy to find which one is preferred. Even more generally, one should allow for an arbitrary distribution of 3-branes along the holographic radial dimension, which might lead to solutions with even lower free energies. Relatedly, we also hope to address more general questions of stability of these solutions, as well as a more detailed look at their physics, including the computation of correlation functions and loop operators. In addition, from a gravitational vantage point, since the shells take over parts of the phase diagram at vanishing temperature, they could play a role in addressing the fate of the weak gravity conjecture \cite{Arkani-Hamed:2006emk,Harlow:2022gzl}; see \cite{Montero:2018fns,Henriksson:2019ifu,McInnes:2022tut} for some recent discussions in the holographic context. 

We expect that it will be possible to find exact, top-down embeddings of our setup, with only minor modifications in, {\emph{e.g.}}, the dual of $\mathcal{N}=4$ SYM at non-zero R-charge density. One must then deal with the complication of the extra internal dimensions. However, in the most symmetric cases, such as when the three chemical potentials of $\mathcal{N}=4$ SYM are set equal, one can perhaps smear the nucleated branes over the internal manifold, thus recovering something close to the effective five-dimensional setup studied here.

One of our motivations for studying finite-density Higgs phases is to model dense nuclear matter. As mentioned in the introduction, the cores of compact stars may hold nuclear matter in such phases, potentially offering experimental access through astrophysical observations. We hope that a better understanding of Higgs phases in holography will offer qualitative insights into the QCD phase diagram, as well as ideas on how to implement them in more realistic phenomenological models \cite{Jarvinen:2021jbd,Hoyos:2021uff}. With this in mind, it is particularly important to study thermodynamics and transport properties of these solutions, since they can be related to astrophysical observables \cite{Hoyos:2020hmq}. 

Finally, let us spell out a feasible and captivating extension of the shell construction. In principle, it could be possible to have flux inside the shells exceeding the value quoted in \eqref{eq:fmax}. A natural interpretation then is that there is some anti-3-brane charge in the interior of the shell that is screened by the 3-branes at the shell. In this case the cosmological constant would be positive in the interior and the spacetime would become de Sitter. This scenario would serve as a simple anchor for studies of quantum gravity in de Sitter spacetime using holography, along the lines of the $dS_2$ inside $AdS_2$ `centaur geometry' proposed in \cite{Anninos:2017hhn}. Extending the centaur geometry to higher dimensional geometries has proved to be a difficult task \cite{Kiritsis:2019wyk}, however, so it would be particularly interesting to have access to a tractable toy model.

\vspace{1cm}

\addcontentsline{toc}{section}{Acknowledgments}
\textit{Acknowledgments }
We would like to thank Ulf Danielsson and Daniel Panizo for discussions on related topics. O.~H. has been supported by the Academy of Finland grant no. 1330346 and by the Ruth and Nils-Erik Stenb\"ack foundation. C.~H. is partially supported by the AEI and the MCIU through the Spanish grant PID2021-123021NB-I00 and by FICYT through the Asturian grant SV-PA-21-AYUD/2021/52177. N.~J. and A.~P. have been supported in part by the Academy of Finland grant no. 1322307. O.~H. and N.~J. acknowledge the hospitality of the APCTP where part of this work was done.

\appendix

\section{D3-brane probe in a spinning black brane background}\label{app:10Dprobemodel}

Many of the choices of the five-dimensional bottom-up model are motivated by earlier top-down computations in \cite{Henriksson:2019zph}. Here we will review the basic results from this model following the above and \cite{Cvetic:1999xp}. The geometry describing the spinning black branes --- dual to $\mathcal{N}=4$ SYM at non-zero R-charge density --- has the metric
\begin{equation}
    ds_{10}^2 = ds_5^2 + L^2 \sum_{i=1}^3 \qty[d\sigma_i^2 + \sigma_i^2 \qty(d\phi_i + L^{-1}A)^2] \ ,
\end{equation}
where $L$ is the radius of curvature, $X_i$ are scalar fields, and $\sigma_i$ satisfy $\sum_i \sigma_i^2 = 1$. Note that the theory in general admits three independent angular momenta (dual to the R-charges); here we set them all equal.

The type IIB superstring equations of motion admit a charged black hole solution described by
\begin{equation}
    ds_5^2 = -H(r)^{-2} f(r) dt^2 + H(r)\qty[\frac{dr^2}{f(r)} + r^2 d\Omega_{3}^2] \ ,
\end{equation}
with the functions
\begin{equation}
    f(r) = 1 - \frac{m}{r^2} + \qty(\frac{r}{L})^2 H(r)\ , \qquad H(r) = 1 + \frac{q^2}{r^2} \ .
\end{equation}
The location of the largest root of $f(r)$ defines the horizon radius $r_H$. The one-form $A$ --- a gauge field from the five-dimensional point of view --- can be written as
\begin{equation}
 A = \frac{q}{r_H^2 + q^2}\sqrt{r_H^2 + q^2 + \frac{r_H^4}{L^2}H(r_H)^3}\qty(1 - \frac{r_H^2 + q^2}{r^2 + q^2})dt \ .
\end{equation}
There is also a four-form Ramond-Ramond potential $C_4=(C_4)_t\, dt\wedge\epsilon_3+\sum_i (C_4)_{\phi_i} d\phi_i\wedge\epsilon_3$, with
\begin{equation}
(C_4)_t = \frac{1}{L}\left[(r^2+q^2)^2 - (r_H^2+q^2)^2\right] \quad \text{and} \quad (C_4)_{\phi_i}= L^2 q \sqrt{ r_H^2 + q^2 + \frac{r_H^4}{L^2} h(r_H)^3}\, \sigma_i^2 \ .
\end{equation}

We can now probe this geometry with a D3-brane; the relevant action is
\begin{equation}
\label{eq:ProbeBraneAction}
    S_{\text{D}3} = S_{\text{DBI}} + S_{\text{WZ}} = -T_3\int d^4 \xi \sqrt{- \det P[G]} + T_3 \int P[C_4] \ ,
\end{equation}
with the same notation as in Section~\ref{sec:model}, except that the DBI-term now contains the 10D metric $G_{\mu\nu}$. We allow the brane to move in the temporal, radial and the three angular directions $\phi_i$ (with equal angular velocity due to the equality of the background angular momenta). It is useful to define $G_{t\phi}=\sum_i G_{t\phi_i}$, $G_{\phi\phi}=\sum_i G_{\phi_i\phi_i}$, and $(C_4)_{\phi}=\sum_i (C_4)_{\phi_i}$. Then, we can write the ten-velocity of the brane as
\begin{equation}
    \dot X = \dot{T}(\tau)\partial_t + \dot{R}(\tau)\partial_r + \sum_{i = 1}^3 \dot{\Phi}(\tau) \partial_{\phi_i} \ ,
\end{equation}
with a dot denoting a $\tau$-derivative, whose square
\begin{equation}
    \dot X_\mu \dot X^\mu = -G_{tt}\dot{T}^2 + G_{rr}\dot{R}^2 + 2G_{t\phi}\dot{\Phi}\dot{T} + G_{\phi\phi} \dot{\Phi}^2
\end{equation}
equals $-1$ if $\tau$ is taken to be proper time. This makes the metric and pullback of the gauge field
\begin{align}
    ds_4^2 & = -\dot X_\mu \dot X^\mu d\tau^2 + (R^2+q^2)d\Omega_3^2 \\
    P[C_4] & = \qty[(C_4)_t \dot{T} + (C_4)_{\phi} \dot{\Phi}]dt \wedge \epsilon_3 \ ,
\end{align}
where $\epsilon_3$ is the volume element of $d\Omega_3^2$, and $\chi_i$ are the spatial coordinates of the brane. The action for a probe brane becomes
\begin{equation}
    S_{\text{D}3} = \int d\tau\wedge\epsilon_3\, \mathcal{L}_{\text{D}3} = -T_3 \int d\tau\wedge\epsilon_3 \qty{(R^2+q^2)^{3/2}\sqrt{-\dot X_\mu \dot X^\mu} - (C_4)_t \dot{T} - (C_4)_{\phi} \dot{\Phi}} \ .
\end{equation}

The action is independent of $\Phi$ (only depending on its derivative), leading to a conserved angular momentum $J$. Legendre transforming to substitute angular velocity for angular momentum gives
\begin{equation}
    \mathcal{L}_{\text{D}3}^J = \mathcal{L}_{\text{D}3} - \dot\Phi J \propto -(C_4)_t  - \frac{J}{L}A_t  + \sqrt{-\left(Z^6 + \frac{J_C^2}{G_{\phi\phi}}\right) \left( G_{tt}-\frac{G_{t\phi}^2}{G_{\phi\phi}} + G_{zz} Z'^2 \right) } \ ,
\end{equation}
where we have used the fact that $G_{t\phi}\sim A_t$. The similarity to our proposed 3-brane action (\ref{eq:braneAction}) is now clear, in particular the term coupling the brane to the one-form $A$.

\bibliographystyle{apsrev4-1}
\bibliography{bibliography}

\end{document}